\documentclass[twocolumn,prl,floatfix,citeautoscript,nofootinbib,superscriptaddress]{revtex4}
\usepackage{amsbsy}
\usepackage{latexsym,epsfig,graphicx}
\usepackage{dcolumn}
\usepackage{graphicx}
\usepackage{subfigure}
\usepackage{comment}
\usepackage{color}
\usepackage{bm}
\usepackage{mathrsfs}
\usepackage{amsfonts}
\usepackage{amsmath}
\usepackage{color}
\usepackage{amssymb}
\usepackage{xspace}
\usepackage{epstopdf}
\usepackage{tabularx}
\usepackage{longtable}
\usepackage[colorlinks=true, letterpaper=true, pdfstartview=FitV, linkcolor=red, citecolor=blue, urlcolor=blue]{hyperref}
\usepackage[normalem]{ulem}

\begin{document}
\title{A continuous variable quantum battery with wireless and remote charging}
\begin{abstract}
Quantum battery has become one of the hot issues at the research frontiers of quantum physics recently. Charging power, ergotropy and wireless charging over long-distance are three important aspects of interest. Non-contact electromagnetic interaction provides an important avenue for wireless charging. In this paper, we design a wireless and remote charging scheme based on the quantized Hamiltonian of two coupled LC circuits and investigate the charging dynamics of a continuous variable quantum battery. It is found that the system can obtain more ergotropy only when the characteristic frequency of the battery is larger than that of the charger. The quantum coherence is more significant than the quantum entanglement for the ergotropy of the quantum battery, regardless of whether the interaction between the charger and the battery is rotating or counter-rotating wave coupling. The feasibility of enabling the battery to extract ergotropy from the thermal reservoir through interplay between the charger and the environment is demonstrated when the roles of the rotating and counter-rotating wave couplings are considered simultaneously. Our wireless charging scheme is not only simple and cost-effective but also offers a longer charging distance than existing qubit batteries.

\end{abstract}

\author{Jun Wen}
\affiliation{School of Mathematics, Physics and Statistics, Sichuan Minzu College, Ganzi 626001, China}
\author{Zheng Wen}
\affiliation{Office of Physics and Chemistry, Army Academy of Border and Coastal Defence, Xi'an 710108, China}
\author{Ping Peng}
\affiliation{School of Physics and Information Science, Shaanxi University of Science and Technology, Xi'an 710021, China}
\author{Guan-Qiang Li}
\thanks{Corresponding author email: liguanqiang@sust.edu.cn}
\affiliation{School of Physics and Information Science, Shaanxi University of Science and Technology, Xi'an 710021, China}

\maketitle

\section{I.~Introduction}
The chemical battery was invented more than $200$ years, and during this time, researchers have continued to boost its performance, and it has now been applied to all aspects of human's lives. With the development of science and technology and the continuous growth of people's living needs, the miniaturization of instruments and equipment has become one of most important development trends. The battery is no exception. When the devices is as small as microns or even nanometers in size, the quantum effect dominates its performance. The concept of the quantum battery is born during the process of the miniaturization. Nowadays, More and more people are looking for ways to apply quantum theories and technologies to battery's research in order to sustainingly improve its performance in all aspects. Although the chemical battery is based on much quantum knowledge, the way the quantum battery works is entirely different~\cite{F.Campaioli2024}.

The fundamental theory of the quantum battery is first and systematically investigated by R. Alicki and M. Fannes~\cite{R.Alicli2013} in $2013$ based on the work given in Ref.~\cite{Allahverdyan2004}. The upper limit of the battery's ergotropy (i.e., maximum extractable work) is estimated by the properties of the free energy and the Gibbs function. It is pointed out that the quantum entanglement in collective charging of quantum batteries is helpful for extracting more ergotropy. Based on this, the research for the quantum battery has made important progress in more than a decade. Progress consists of two main aspects: one is to provide proposals for new schemes and systems that can be used to create quantum batteries; the other is to study the impact of quantum resources on the performance of quantum batteries, focusing on ergotropy and charging power. The new proposals included, but were not limited to, nonreciprocal~\cite{BAhmadi2024} and topological~\cite{XYLu2025} quantum batteries. The charging scheme with wireless energy transfer was also proposed recently~\cite{HuML2025, Dou2025}. The systems for supporting the quantum batteries have been extended to ultracold atomic gases~\cite{TKKonar2022}, miniature microwave RFs~\cite{Shaghaghi2022, Rodriguez2023}, and nuclear spins~\cite{J.Joshi2022}. The effects of the relevant parameters of these systems on the performance of the batteries were investigated correspondingly. As a unique resource for quantum systems, the entanglement has become the focus of research. A subset of researchers supports the positive impact of the entanglement on quantum batteries~\cite{K.V.Hovhannisyan2013, F.C.Binder2015, D.Ferraro2018, D.Rosa2020}. However, other researchers show that the entanglement can also have negative effects~\cite{FLiu2025, J.X.Liu2021, JWen2025}. The entanglement of the quantum battery based on the Dicke model was revealed to benefit the coherent work but suppress the incoherent work, and it is pointed out that either the entanglement between the battery and the charger or the quantum coherence inside the battery is important for generating ergotropy~\cite{H.L.Shi2022}. Actually, whether the entanglement is beneficial or detrimental to quantum batteries depends on the specifics~\cite{JYGyhm2024, SYLiu2023}. Meanwhile, the positive role of quantum coherence as another quantum resource for a battery has been confirmed~\cite{STirone2025, FHKamin2020}. Some research even considers the environment interacting with the system as a resource for charging quantum battery. In $2021$, the spin-chain quantum battery in an optical cavity coupled to a Markovian thermal reservoir was investigated and it was shown that the ergotropy can be extracted during the thermal charging~\cite{F.Zhao2021}. Recently, the charging system including two qubits as quantum battery and charger was studied so that a long-distance wireless charging could be achieved by coupling a shared non-Markovian environment~\cite{Song2024}. Using the parameters given in that paper, it can be estimated that the ergotropy of the quantum battery tends to be zero when the distance between the two qubits is greater than 1.5$\mu m$. The above studies are about the qubit batteries and have provided valuable data for realistic applications of quantum batteries.

In recent years, more and more attention has been paid to quantum batteries based on continuous variables. In $2023$, a method was proposed for improving the intrinsic charging performance and energy storage of quantum batteries by incorporating a catalyst system (which can be a qubit or a quantum resonator) between the battery and the charger~\cite{Roddriguez2023}. Numerical simulation showed that the catalyst system does not require energy expenditure. In the same year, a continuous variable quantum battery by preparing the initial state of the charger as a Gaussian state was investigated under the conditions of considering the weak coupling between the battery and the charger and exposing the charger to a Markovian environment. The relationship between the ergotropy of the battery and the dissipation/coupling coefficients was analyzed, and the effect of the dissipation and coupling coefficients on the quantum phase transition of the system was investigated~\cite{Downing2023}. In $2024$, a driving term $\Omega \delta (t)(\hat{a}^{\dagger}\hat{a}^{\dagger}+\hat{a}\hat{a})/2$ was used to investigate the performance of a continuous variable quantum battery~\cite{Downing2024}. It was found that this driving term not only hyperbolically increases the energy stored in the battery, but also suppresses the dissipation effect. In the same year, an approach was proposed to achieve a unidirectional flow of energy from the charger to the quantum battery~\cite{BAhmadi2024}. In the case of an open system, the goal of extracting the ergotropy in the quantum battery to be a nonzero constant after a sufficiently long period of time was achieved. The approach also increases the energy storage capacity of the battery and eliminates the need for disconnecting the coupling between the battery and the charger at a precise point in time. Clearly, the continuous variable quantum batteries have huge potentials for extracting useful work.

It is well known that the mutual induction between electromagnetic fields has been widely used in many fields. In quantum optics, the alternating electromagnetic fields or LC circuits can be quantized as resonators~\cite{Scully2012, A.Blasi2021, Q.Xu2022}, and the wireless and remote charging can be achieved by coupling the capacitance and inductance of two LC circuits. Inspired by these theories, we have some questions: What will be the role of quantum resources such as entanglement and coherence for the continuous variable quantum batteries? In the case of wireless charging, can the distance between the quantum battery and the charger be extended even further? It has been established that the thermal charging processes of single-mode quantum batteriy does not allow for the attainment of nonzero ergotropy~\cite{D.Farina2019}. Therefore, it is pertinent to inquire how the thermal charging of a single-mode continuous variable quantum battery would differ. In this paper, we theoretically propose the implementation of a charging scheme for the continuous variable quantum batteries, and investigate the charging dynamics of: (i) a closed system consisting of a battery and charger when the rotating or counter-rotating wave coupling exists alone; (ii) a closed system consisting of a battery and charger when both of the rotating and counter-rotating wave couplings exist simultaneously; (iii) an open system consisting of a battery, charger, and Markovian environment when both of the rotating and counter-rotating wave couplings are present. By theoretical analysis and numerical calculation of the equations describing the closed and open quantum systems, the answers for the above questions will be given step by step in the following context.

\section{II.~Model and solving method }
\begin{figure}[tbph]
\centering\includegraphics[width=7.5cm]{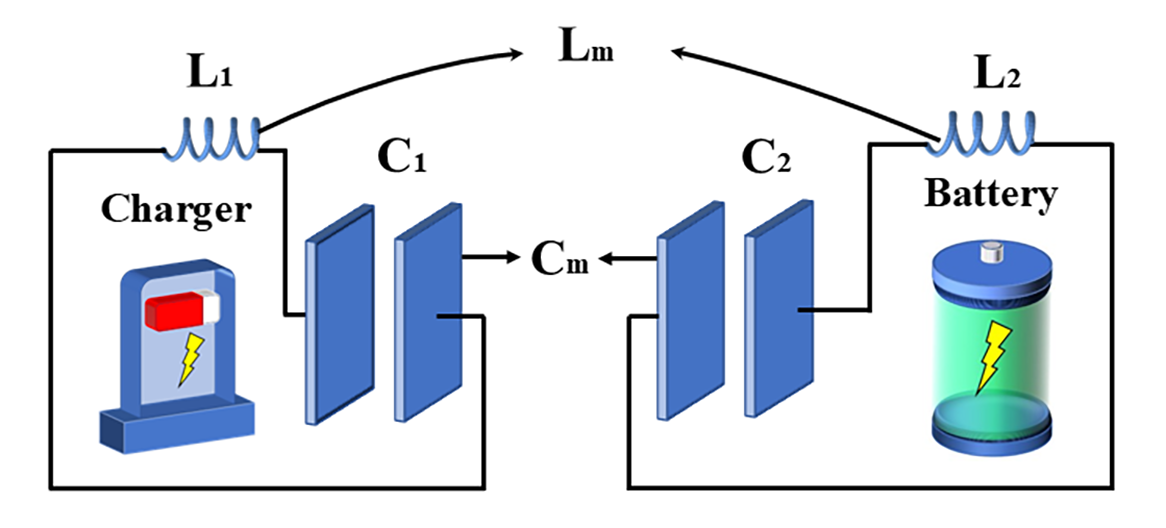}
\caption{(Color online) Scheme of a continuous variable quantum battery with wireless and remote charging based on two coupled LC circuits. $L_{1,2}$ $(C_{1,2})$ denote the inductances (capacitances) of the two LC circuits. The $L_{m}$ $(C_{m})$ is the coupling inductance (capacitance) between the circuits. The coupling strength is determined by the relative position and distance of the inductors and capacitors. }
\label{figure1}
\end{figure}
The scheme of a continuous variable quantum battery with wireless and remote charging based on two coupled LC circuits is shown in Fig.~\ref{figure1}. The coupling between the two LC circuits include the contribution from both of the inductors and the capacitors. The energy between the two inductors can be expressed as
\begin{equation}
H_{L}=\frac{1}{2}I_{1}\Phi _{1}+\frac{1}{2}I_{2}\Phi _{2},  \label{equation-1}
\end{equation}%
where $I_{j}$ and $\Phi _{j}$ are the current and magnetic flux of the $j$th ($j=1, 2$) coil, respectively. Although this formula looks like only considering the energy of each inductor, it actually includes the energy of the interaction. The magnetic flux in each coil consists of two parts: one part of the flux is excited by the coil's own current and the other part is generated by the current in the other coil. That is
\begin{equation}
\Phi _{1}=L_{1}I_{1}+L_{m}I_{2},~~~\Phi _{2}=L_{2}I_{2}+L_{m}I_{1}, \label{equation-2}
\end{equation}%
where $L_{m}$ is the mutual inductance, $k_{L}=L_{m}/\sqrt{L_{1}L_{2}}$ is
the magnetic coupling coefficient. By combing Eq.~(\ref{equation-1}) and Eqs.~(\ref{equation-2}), the magnetic energy stored in the two coils
\begin{equation}
H_{L}=\frac{L_{1}\Phi _{2}^{2}}{2(L_{1}L_{2}-L_{m}^{2})}+\frac{L_{2}\Phi
_{1}^{2}}{2(L_{1}L_{2}-L_{m}^{2})}-\frac{L_{m}\Phi _{1}\Phi _{2}}{%
L_{1}L_{2}-L_{m}^{2}}.  \label{equation-3}
\end{equation}%
Similar with the derivation process of the magnetic energy, the energy in the two capacitors is
\begin{equation}
H_{C}=\frac{C_{1}Q_{2}^{2}}{2(C_{1}C_{2}-C_{m}^{2})}+\frac{C_{2}Q_{1}^{2}}{%
2(C_{1}C_{2}-C_{m}^{2})}-\frac{C_{m}Q_{1}Q_{2}}{C_{1}C_{2}-C_{m}^{2}},
\label{equation-4}
\end{equation}%
where $C_{j}$, $Q_{j}$ and $C_{m}$ denote the capacitance, charge, and coupling capacitance between the capacitors, and the electric coupling coefficient is $k_{C}=C_{m}/\sqrt{C_{1}C_{2}}$~\cite{L.Huang2015}. As outlined in Refs.~\cite{Scully2012, A.Blasi2021, Q.Xu2022}, the total energy of the system is $H=H_{C}+H_{L}$.

The classical electromagnetic energy can be transformed into the quantized Hamiltonian:
\begin{eqnarray}
\widehat{H} &=&\widehat{H}_{0}+\widehat{H}_{i},~~~~ \label{equation-5}   \\
\widehat{H}_{0} &=&\hbar \omega _{1}\hat{a}_{1}^{\dagger }\hat{a}_{1}+\hbar \omega _{2}%
\hat{a}_{2}^{\dagger }\hat{a}_{2},~~~~  \notag \\
\widehat{H}_{i} &=&g\hbar (\hat{a}_{1}^{\dagger }+\hat{a}_{1})(\hat{a}_{2}^{\dagger }+%
\hat{a}_{2})+G\hbar (\hat{a}_{1}^{\dagger }-\hat{a}_{1})(\hat{a}%
_{2}^{\dagger }-\hat{a}_{2}),~~~~   \notag
\end{eqnarray}
where $\omega _{j}=[ L_{j}C_{j}(1-k_{C}^{2})(1-k_{L}^{2})]^{-\frac{1}{2}}$ represents the characteristic frequency of the $j$th LC circuit. $\widehat{H}_{0}$ denotes the sum of the energies for the charger and the battery, $\widehat{H}_{i}$ corresponds to the interaction energy between the charger and the battery. The coupling coefficients in the interacting Hamiltonian are $g=-k_{L}\sqrt{\omega _{1}\omega_{2}}/2$ and $G=k_{C}\sqrt{\omega _{1}\omega _{2}}/2$. According to the classical electromagnetic theory, the values of $k_{C}$ and $k_{L}$ belongs to the range of $(-1, 1)$, and the specific values is determined by the relative position of the inductor and capacitor as well as the distance between them. The system only includes the counter-rotating wave coupling if $G=g$ and the rotating one if $G=-g$. The relative sizes of the parameters $\omega _{1}$, $\omega_{2}$, $G$ and $g$ determine the system's charging dynamics. In the subsequent discussion, the first (second) circuit is taken as the charger (battery), with the corresponding ascending and descending operators denoted by $\hat{a}_{1}^{\dagger}$ ($\hat{a}_{2}^{\dagger}$) and $\hat{a}_{1}$ ($\hat{a}_{2}$), as shown in Fig.~\ref{figure1}.

We introduce the physical quantities for characterizing the charging dynamics of our battery system as follows: $E_{1}(t)$, $E_{2}(t)$, $E_{e}(t)$ and $E_{i}(t)$ denote the energy of the charger, the energy stored in the battery, the ergotropy, and the interaction energy which is given by the expectation value of $\widehat{H}_{i}$; $R(t)$, $S(t)$ and $C(t)$ denote the ratio $E_{e}(t)/E_{2}(t)$, the Von Neumann entropy and the quantum coherence of the battery, respectively. Using these quantities, not only the performance of the battery but also the effect of the quantum entanglement and coherence on the battery can be analyzed. The algorithms for the corresponding energies are given as $E_{1}(t)=\hbar\omega _{1}\mathrm{Tr}[\hat{\varrho} _{1}(t)\hat{a}_{1}^{\dagger}\hat{a}_{1}] $, $E_{2}(t)=\hbar\omega _{2}\mathrm{Tr}[\hat{\varrho} _{2}(t)\hat{a}_{2}^{\dagger}\hat{a}_{2}]$ and $E_{e}(t)=E_{2}(t)-\sum\nolimits_{k}r_{k}e_{k}$, where $\hat{\varrho}_{j}$ denotes the reduced density matrix of the $j$th subsystem. $r_{k}$ is the $k$th eigenvalue of the battery's Hamiltonian $\hbar\omega _{2}\hat{a}_{2}^{\dagger }\hat{a}_{2}$ in ascending order and $e_{k}$ is the $k$th eigenvalue of $\varrho_{2}(t)$ in descending order~\cite{Allahverdyan2004}. $R(t)$ gives the proportion of the ergotropy in the battery's energy. The entropy $S(t)$ describes the disorder of the subsystem and the degree of entanglement of the whole system. The initial state of the quantum battery is set to be the vacuum state, i.e., the Fock state with $n=0$. The state of the charger is initially set to be the coherent or thermal state. The calculation method of the quantum coherence $C(t)$ for the Gaussian state was given in Ref.~\cite{J.W.Xu2016}. The Gaussian state can be diagonalized analytically, so $\sum\nolimits_{k}r_{k}e_{k}=\hbar\omega _{2}(\sqrt{D}-1)/2$, where $D$ is the determinant of the covariance matrix of the Gaussian state~\cite{Weedbrook2012}.

The dynamics of the Gaussian states is obtained by solving $\frac{d\langle\widehat{O}\rangle}{dt}=\mathrm{Tr}\left[\widehat{O}\frac{d\hat{\varrho} (t)}{dt}\right]$~\cite{Carmichael1989}. $\widehat{O}$ is an arbitrary time-independent operator and $\langle \widehat{O}\rangle$ denotes its expectation value. The derivative of the density matrix $\hat{\varrho}(t)$ with respect to time is determined by the master equation of the system. The equation is written as follows~\cite{Scully2012}:
\begin{eqnarray}
\frac{d\hat{\varrho}(t)}{dt}&=&\frac{i}{\hbar}[\hat{\varrho} (t),\widehat{H}]  \notag \\
&& +\frac{\gamma n_{th}}{2}[2\hat{a}_{1}^{\dag}\hat{\varrho}(t)\hat{a}_{1}-\hat{a}_{1}\hat{a}_{1}^{\dag}\hat{\varrho}
(t)-\hat{\varrho}(t)\hat{a}_{1}\hat{a}_{1}^{\dag}]  \notag \\
&& +\frac{\gamma (n_{th}+1)}{2}[2\hat{a}_{1}\hat{\varrho}
(t)\hat{a}_{1}^{\dag}-\hat{a}_{1}^{\dag}\hat{a}_{1}\hat{\varrho}(t)-\hat{\varrho}(t)\hat{a}_{1}^{\dag}\hat{a}_{1}],  \label{equation-6} \notag \\
\end{eqnarray}
where $\gamma$ denotes the decay coefficient induced by the environment (i.e., the thermal reservoir). For the closed systems, the coefficient need to be set to zero $(\gamma=0)$. For the Markovian open systems, the coefficient is a nonzero constant $(\gamma\neq0)$. $n_{th}$ is the mean photon number for the mode in the thermal reservoir resonating with the charger's frequency. In Eq.~(\ref{equation-6}), the battery is isolated from the environment for avoiding the decay of the quantum coherence and the thermalization, but the interaction between the charger and the environment is considered for effectively extracting the thermal energy~\cite{F.Zhao2021, Downing2023}.

Noting that the analytical solution of the charging dynamics can be obtained by the Laplace transform of the Heisenberg equations for the closed system governed by the Hamiltonian (\ref{equation-5}) (which involves in solving the quartic algebraic equation with single variable). For studying the open system, it is necessary to solve the master equation (\ref{equation-6}) numerically. The coupling between the battery and the charger involves only the quadratic power of the ascending and descending operators. As long as the initial quantum state is the Gaussian state, the subsequent quantum states will keep in the Gaussian state~\cite{A.Serafini2017}. The Gaussian state can be described by its covariance matrix and $\langle\hat{a}_{2}^{\dagger}\pm \hat{a}_{2}\rangle$. If the solutions of the corresponding equations for the closed and open systems are obtained, all of the information about the quantum battery system can be determined [See the derivation and solving methods of the equations in the Appendix].
\section{III.~Main results}
\subsection{A.~For the closed system with rotating or counter-rotating wave coupling}
\begin{figure}[tbph]
\centering\includegraphics[width=10 cm]{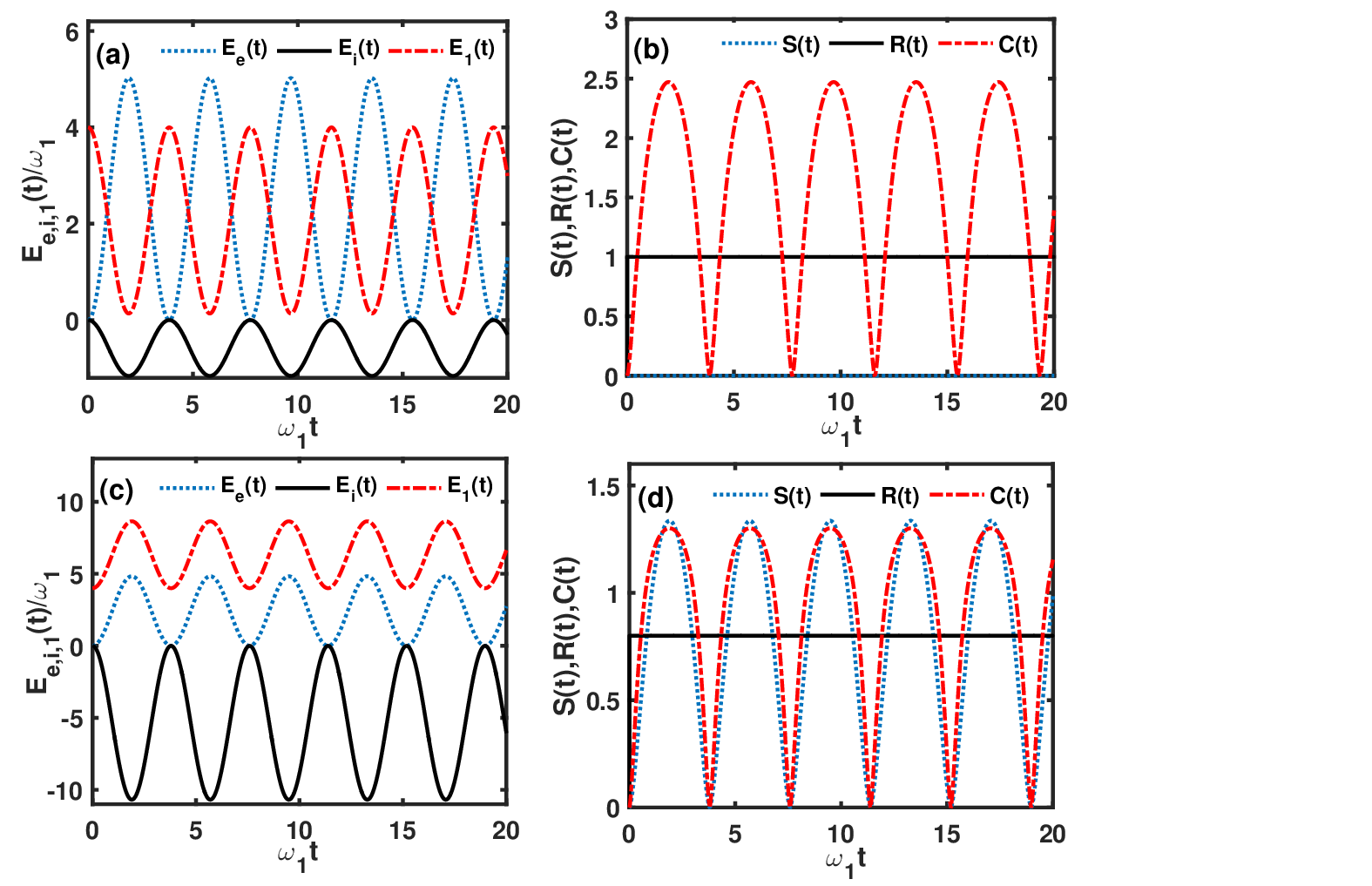}
\caption{(Color online) Change of the physical quantities with time for the closed system. (a, c) demonstrate the dynamics of $E_{e}(t)$, $E_{i}(t)$ and $E_{1}(t)$, while (b, d) give the dynamics of $S(t)$, $R(t)$ and $C(t)$. The system contains only the rotating wave coupling for $k_{L}=k_{C}=0.7$ in (a) and (b) and only the counter-rotating wave coupling for $k_{L}=-k_{C}=-0.7$ in (c) and (d). The coherent state $\protect\hat{\varrho}_{1}(0)=\left\vert \protect\alpha \right\rangle\left\langle\protect\alpha\right\vert$ with $\protect\alpha=2$ is chosen as the initial state of the charger and $\omega _{2}=1.3\omega _{1}$. }
\label{figure2}
\end{figure}

Most of recent research has investigated the qubit battery in which the frequency of the battery resonates with that of the charger to ensure the maximum energy transfer~\cite{Downing2023, Downing2024, FZhao2021}. By comparison, the continuous variable quantum battery discussed in our model exhibit four new features. The first feature is that the more ergotropy can be obtained when the frequency of the battery is larger than that of the charger $(\omega_{2}>\omega_{1})$. The total energy of the system is defined as zero, with the charger and the battery both in their respective vacuum states. For our model, the maximum value of $E_{e}(t)$ is $4\hbar\omega_{1}$ for $\omega _{1}=\omega_{2}$ under only considering the rotating wave coupling when the initial state of the charger is the coherent state $\protect\hat{\varrho}_{1}(0)=\left\vert \protect\alpha \right\rangle\left\langle\protect\alpha\right\vert$ with $\protect\alpha=2$. The interaction energy $E_{i}(t)$ keeps zero in the whole process of the evolution because there only exists the energy exchange between the battery and the charger. Fig.~\ref{figure2}(a) shows the evolution of the energies $E_{1}(t)$, $E_{e}(t)$ and $E_{i}(t)$ with time for $\omega_{2}=1.3\omega_{1}$ when the rotating wave coupling is considered (i.e., $G=-g$). The oscillation of $E_{e}(t)$ is completely out of phase with $E_{1}(t)$ and $E_{i}(t)$. The negative interaction energy $E_{i}(t)$ demonstrates the attractive coupling between the charger and the battery. For $\omega_{2}>\omega_{1}$, the battery can acquire more energy from the system. Since the energy of the system is conserved, the extra energy is derived from the part of the reduction in the interaction energy. One photon from the charger cannot produce one photon in the battery, resulting in a part of the interaction energy is stored in the battery. The larger the battery's frequency, the more energy can be extracted from the system. But, it does not mean that the battery can obtain an infinite amount of energy if we increase $\omega _{2}$ continuously when the charger's frequency $\omega _{1}$ is fixed. In fact, there is an upper limit for $E_{e}(t)$. Specifically, the limit is about $7.84\hbar\omega _{1}$ for the parameters given in Fig.~\ref{figure2}(a) except for $\omega_{2}=10^{4}\omega _{1}$. Furthermore, for the system only considering the rotating wave coupling, above $80\%$ of the maximum energy for the battery does not come from the charger but the interaction energy when $\omega _{2}=6\omega _{1}$ is satisfied. The evolution of the battery is also faster than that of the charger for $\omega_{2}>\omega_{1}$. The enhancement of charging power (i.e., the derivative of $E_{2}(t)$ with time) can be realized by increasing $\omega _{2}$, since the increase of the coupling strengthes $|g|$ and $|G|$ accelerates the energy transferring from the charger to the battery. For $\omega_{2}<\omega_{1}$, the battery's energy is less than $4\hbar\omega _{1}$. The evolution of the charger is faster than that of the battery, a portion of the charger's energy is stored as the interaction energy and the energy of the battery is decreased.

Fig.~\ref{figure2}(b) shows the evolution of $R(t)$, $S(t)$ and $C(t)$ with time for only considering the rotating wave coupling. All of the energy stored in the battery can be fully used due to $R(t)=1$. It means that the determinant of the covariance matrix of the Gaussian state $D=1$ and the quantum state of the battery is pure in the whole process of the evolution~\cite{J.W.Xu2016}. The quantum pure state cannot be transformed into the mixed state by the rotating wave coupling. The quantum coherence $C(t)$ oscillates with the frequency of $2\pi\omega_{1}/3.86$, which is synchronized with the oscillation of the ergotropy $E_{e}(t)$ in Fig.~\ref{figure2}(a). The entropy $S(t)=0$ shows that the entanglement between the charger and the battery is not a prerequisite for achieving a nonzero ergotropy for our system. The dynamics of the ergotropy is determined by the quantum coherence, the quantum entanglement play no role under only considering the rotating wave coupling.

Fig.~\ref{figure2}(c) shows the evolution of the energies $E_{1}(t)$, $E_{e}(t)$ and $E_{i}(t)$ with time for only considering the counter-rotating wave coupling (i.e., $G=g$). It is found that the oscillation of $E_{e}(t)$ is in phase with $E_{1}(t)$ but completely out of phase with $E_{i}(t)$. The energy of the charger $E_{1}(t)$ is always larger than $4\omega_{1}$ in this situation. The battery's energy comes entirely from the interaction, rather than from the charger. The counter-rotating wave coupling inhibits energy exchange between the charger and the battery. The reason is that the existence of the counter-rotating wave coupling makes it difficult to synchronize the dynamical phases of different single-particle states, which can not form an effective enhancement for the system's coherence. It seems that what state the charger is in doesn't matter anymore for this situation. There is a hypothesis that the charger can be directly prepared into the vacuum state, making the battery continuously extract the energy from the interaction. However, this hypothesis does not apply to reality. When the initial energy of the charger is zero, although the system can evolve and the battery can gain the energy, but the latter cannot be transferred into the ergotropy. Fig.~\ref{figure2}(d) demonstrates the evolution of $R(t)$, $S(t)$ and $C(t)$ with time for only considering the counter-rotating wave coupling. $R(t)=0.8$ means about eighty percent of the battery's energy can be extracted throughout the process. The pure state is changed into the mixed one by the counter-rotating wave coupling, and the disorder exists in the system. The presence of nonzero entropy is not conducive to transferring the battery's energy into the ergotropy. The entropy $S(t)$ oscillates over time, and the oscillation of $S(t)$ is synchronized with $C(t)$. A comparison of Fig.~\ref{figure2}(b) with Fig.~\ref{figure2}(d) shows that the oscillation of the ergotropy is consistent with that of the quantum coherence, but not always with that of the quantum entanglement. The second feature that can be deduced is that the quantum coherence is more significant than the quantum entanglement for the ergotropy of the continuous variable quantum battery, irrespective of the wave coupling being rotating or counter-rotating.

\subsection{B.~For the closed system with rotating and counter-rotating wave couplings}
One of the significant advantages of the thermal state is its low cost of preparation. Studying the acquisition of ergotropy from the thermal state is a very meaningful task. For our closed system $(\gamma=0)$, if the battery is prepared in the thermal state, there is no ergotropy within the battery. The reason is that there is no quantum coherence for the thermal state. However, if the initial state of the charger is a thermal state, it is possible to obtain nonzero ergotropy for the battery after charging. The prerequisite is that the interaction between the charger and the battery must contain both the rotating and counter-rotating wave couplings. If only the rotating or counter-rotating wave coupling is included in the interaction, there is no ergotropy in the battery at the same condition. Correspondingly, there is no quantum coherence, implying that the compression effect of the counter-rotating wave coupling is insufficient to produce the ergotropy even the quantum entanglement exists in the system~\cite{Barzanjeh2022}. As mentioned above, the dynamical behaviors for $S(t)$, $R(t)$ and $C(t)$ in Fig.~\ref{figure2}(b) and Fig.~\ref{figure2}(d) are very different. The different quantum states for the battery are induced by the counter-rotating and rotating wave couplings for the same initial states and the same parameters of the system in the process of the evolution. If both of the counter-rotating and rotating wave couplings are considered, the quantum state of the battery will be the superposition of such different quantum states, and the quantum coherence will be nonzero, leading to the nonzero ergotropy~\cite{J.X.Liu2021}.

\begin{figure}[tbph]
\centering\includegraphics[width=10 cm]{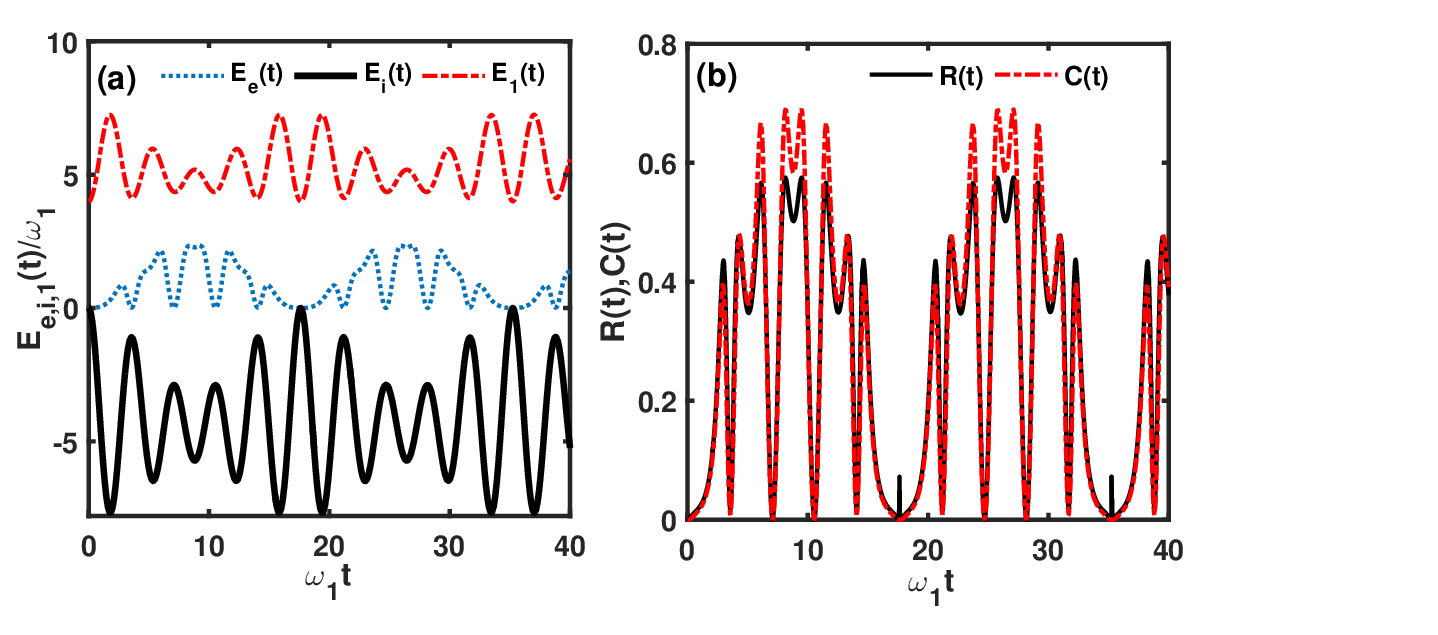}
\caption{(Color online) (a) Dynamics of $E_{e}(t)$, $E_{i}(t)$ and $E_{1}(t)$; (b) Dynamics of $R(t)$ and $C(t)$. The thermal state with an average thermal photon number $n_{p}=4$ is chosen as the initial state of the charger. The other parameters are $\protect\omega _{2}=1.3\omega_{1}$, $k_{L}=-0.57$ and $k_{C}=0.7$. }
\label{figure3}
\end{figure}

Fig.~\ref{figure3} gives the result of the thermal charging for $\gamma=0$ when both of the counter-rotating and rotating wave couplings are considered. Fig.~\ref{figure3}(a) demonstrates the evolution of the energies $E_{1}(t)$, $E_{e}(t)$ and $E_{i}(t)$ with time. It is shown that $E_{1}(t)$, $E_{e}(t)$ and $E_{i}(t)$ oscillate over time and exhibit beat phenomenon. The value of $E_{1}(t)$ is larger than $4\omega _{1}$ for the thermal photon number $n_{p}=4$. The energy of the battery originates from the interaction energy rather than the charger. The interaction energy can be subdivided into two parts, which are related to the rotating and counter-rotating wave couplings. Correspondingly, there exist two effective frequencies corresponding to the couplings in the system. The competition between the couplings leads to the beat phenomenon of the charging dynamics of the quantum battery. The oscillating frequency for the slow component is one half of the difference of the two frequencies (i.e., beat frequency), while the oscillating frequency of the fast component is one half of the sum of the two frequencies. The values for the difference and sum frequencies are about $4\pi\omega_{1}/17.63$ and $4\pi\omega_{1}/3.61$, respectively. Furthermore, the ergotropy $E_{e}(t)$ is nonzero for most of the evolution time. The nonzero ergotropy is related with the nonzero quantum coherence. Fig.~\ref{figure3}(b) demonstrates the evolution of $R(t)$ and $C(t)$ with time. Except the same oscillating behavior, $R(t)$ is synchronized with $C(t)$. $R(t)$ and $C(t)$ are strongly correlated each other. The importance of the quantum coherence is further verified since the nonzero coherence leads to the nonzero ergotropy. Fig.~\ref{figure3} indicates that the utilization of the rotating wave coupling in conjunction with the counter-rotating wave coupling can yield unexpected outcome, i.e., the ergotropy can be extracted for continuous variable quantum battery by the thermal charging, which is the third feature of the quantum battery proposed in the present paper.

\begin{figure}[tbph]
\centering\includegraphics[width=12 cm]{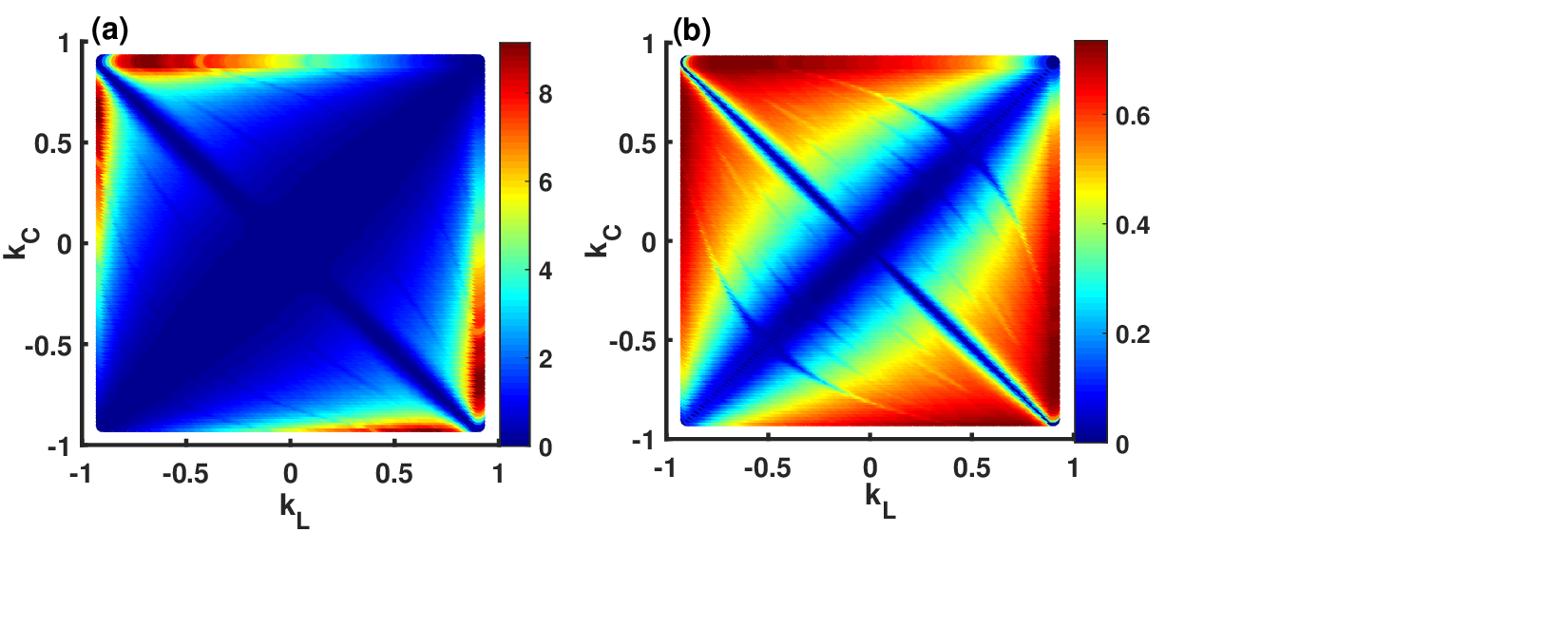}
\caption{(Color online) Max$[E_{e}(t)]$ varies with $k_{L}$ and $k_{C}$ in (a), while Max$[R(t)]$ varies with $k_{L}$ and $k_{C}$ in (b). The thermal state with an average thermal photon number $n_{p}=4$ is chosen as the initial state of the charger. The other parameters are $\protect\omega _{1}t\in \lbrack 0,20]$ and $\protect\omega_{2}=1.3\omega_{1}$. }
\label{figure4}
\end{figure}

\begin{figure}[tbph]
\centering\includegraphics[width=12 cm]{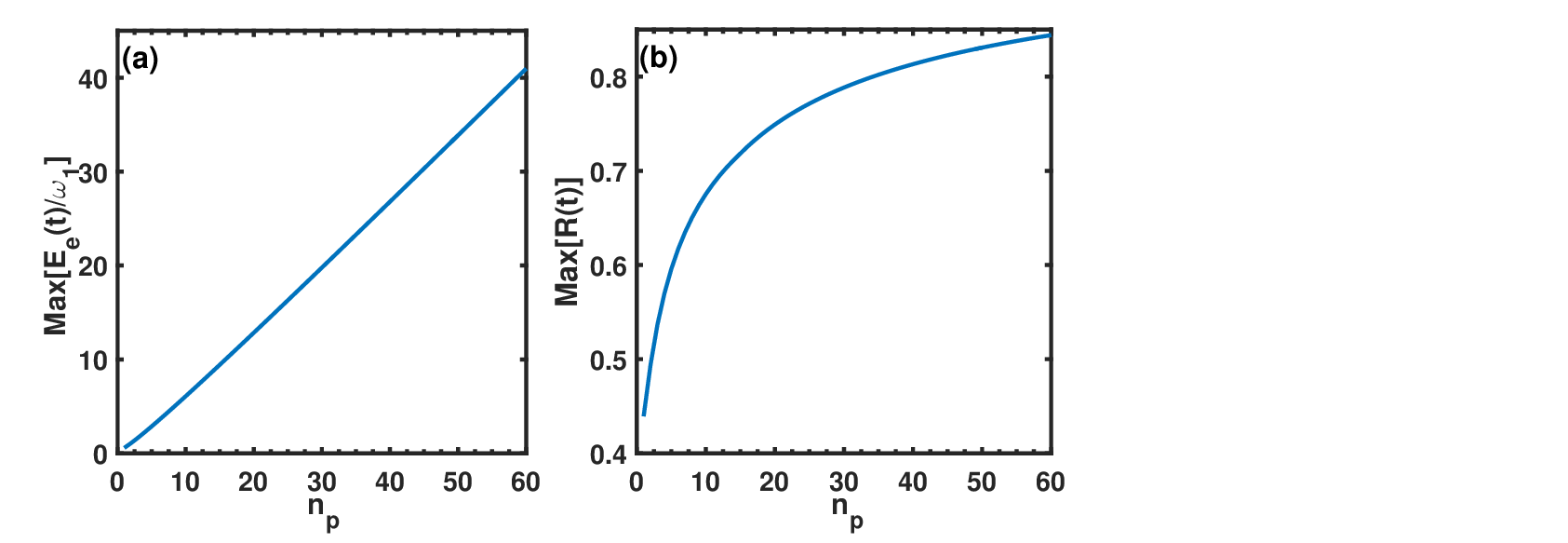}
\caption{(Color online) Change of Max$[E_{e}(t)]$ in (a) and Max$[R(t)]$ in (b) vary with the average thermal photon number $n_{p}$ at $\protect\omega_{1}t\in \lbrack 0,20]$. The thermal state is chosen as the initial state of the charger. The other parameters are $\protect\omega_{2}=1.3\omega_{1}$, $k_{L}=-0.37$ and $k_{C}=0.7$. }
\label{figure5}
\end{figure}

When combining the rotating and counter-rotating wave couplings, what value of $k_{C}$ and $k_{L}$ can be taken to ensure that the ergotropy is maximized for a given set of conditions? Fig.~\ref{figure4} gives the change of Max$[E_{e}(t)]$ and Max$[R(t)]$ with $k_{C}$ and $k_{L}$. It is found that the larger values of Max$[E_{e}(t)]$ and Max$[R(t)]$ always satisfy the condition $k_{C}k_{L}<0$, which indicates that the strength of the counter-rotating coupling is greater than that of the rotating wave coupling (i.e., $|g+G|>|g-G|\geq0$). This can be deduced by means of reasoning based on the conditions of the rotating wave approximation. The counter-rotating wave coupling only plays key role when its strength is substantial, thereby significantly impacts the system's dynamical behaviors. The relationships between the maximum values of $E_{e}(t)$ and $R(t)$ with $k_{C}$ and $k_{L}$ under condition of $n_{p}=4$ are demonstrated in Fig.~\ref{figure4}(a) and Fig.~\ref{figure4}(b), with the aim of providing a reference for the practical operation of the quantum batteries. Indeed, the greater the initial energy, the more power is used to drive the evolution. The maximum ergotropy is monotonically increased by enhancing the average number of photons for a specific evolution period, as shown in Fig.~\ref{figure5}(a). However, it should be noted that the relationship between the maximum value of $R(t)$ and the average number of photons $n_{p}$ over an evolution period cannot be determined by theoretical analysis. It can only be obtained through numerical calculation. Fig.~\ref{figure5}(b) shows that the change of Max$[R(t)]$ with $n_{p}$ is a convex function. So, can the value of Max$[R(t)]$ saturate and approach to $1$? It is predicted that Max$[R(t)]\rightarrow1$ if $n_{p}\rightarrow+\infty$. In order to verify this prediction, $n_{p}=10^{5}$ is taken in the numerical calculation and Max$[R(t)]=0.996$ is realized. In the present paper, we do not study the situation of $|k_{C,L}|\geq1$, which would make the system to enter the region of deep strong coupling and lead to the energy divergence.


\subsection{C.~For the open system with rotating and counter-rotating wave couplings}
\begin{figure}[tbph]
\centering\includegraphics[width=10 cm]{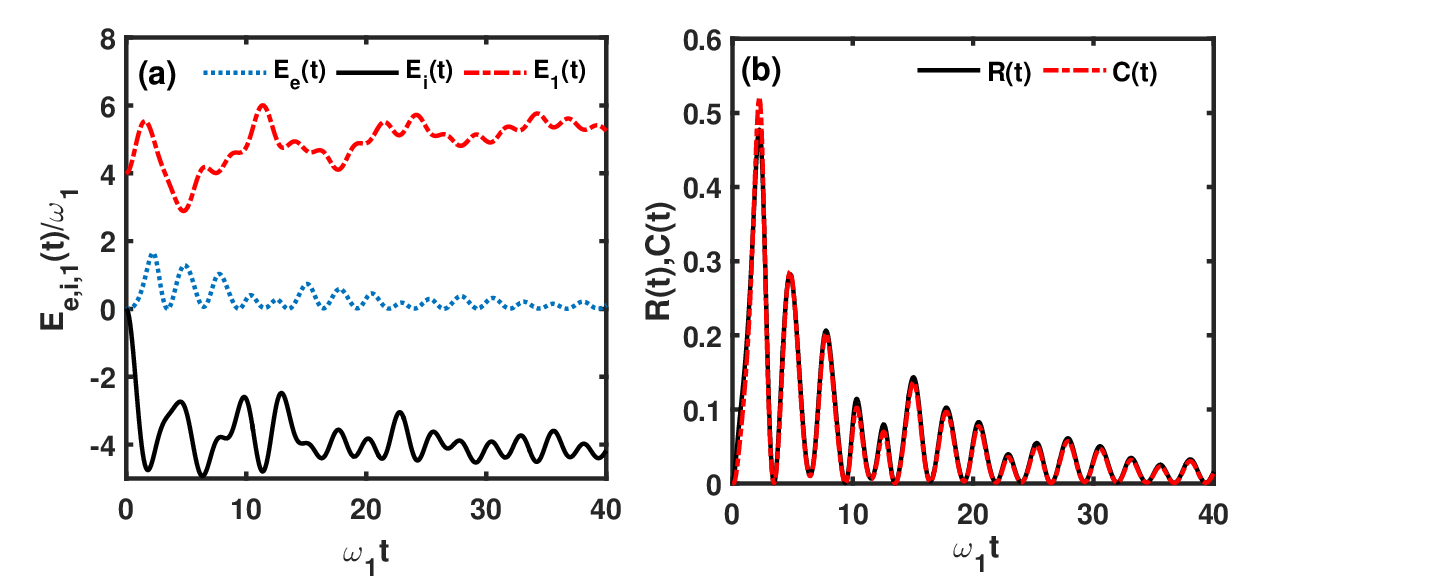}
\caption{(Color online) (a) Dynamics of $E_{e}(t)$, $E_{i}(t)$ and $E_{1}(t)$; (b) Dynamics of $R(t)$ and $C(t)$. The thermal state with the mean number of the thermal photon $n_{p}=4$ is chosen as the initial state of the charger. The photon number corresponding to the mode with the frequency $\omega_{1}$ in the thermal reservoir is chosen as $n_{th}=4$. The other parameters are $\protect\gamma =0.1\omega_{1}$, $\protect\omega_{2}=1.3\omega_{1}$, $k_{L}=-0.37$ and $k_{C}=0.7$. }
\label{figure6}
\end{figure}

If the temperature of the environment is nonzero, the final state of the charger will be a thermal state with nonzero photon number. Since the charger is embedded in the Markovian environment, whether the battery can obtain the ergotropy is another interesting question. Fig.~\ref{figure6} gives the result of the thermal charging for $\gamma\neq0$. As shown in Fig.~\ref{figure6}(a), the ergotropy of the battery will become zero after enough evolution time. The quantum coherence, supported by the competition between the counter-rotating and rotating wave couplings, may dissipate through the environment [Fig.~\ref{figure6}(b)]. Fig.~\ref{figure6} does not given information about the quantum entanglement. The entanglement can no longer be described by the von Neumann entropy because the overall quantum state is a mixed state for our open system. But the entanglement does exist due to the appearance of the counter-rotating wave coupling~\cite{Barzanjeh2022}. At least one of the quantum coherence and entanglement exists in order to generate the ergotropy~\cite{H.L.Shi2022}.

For fully absorbing the energy in the environment, the quantum battery must be strongly coupled with the charger and the latter must be stayed in a Markovian environment. The battery can obtain the ergotropy before the quantum coherence and entanglement have seriously been decayed. The condition $\gamma\ll\min [\left\vert G-g\right\vert ,\left\vert G+g\right\vert ]$ ensures that the quantum coherence and entanglement can not rapidly decay into the environment during the charging period, which buys time for producing the ergotropy. It is possible to achieve LC circuits with frequencies up to $10^{10}$Hz and the coefficient of the decoherence $10^{3}$Hz in current experiments~\cite{Q.Xu2022}. It takes only a few milliseconds or even a dozen milliseconds for the charger to fully absorb the thermal photons in the environment and to reach a stable state. One issue needs to be clarified is that the decay coefficient of $\gamma=0.1\omega_{1}$ in Fig.~\ref{figure6} is far larger than that in the realistic experiments. In fact, we also study the evolution pattern for the case of a time period of $\omega _{1}t\in \lbrack 0,400]$ under $\gamma =0.01\omega_{1}$. It is found that the decay of the oscillations becomes slower but the general pattern of the evolution is same with that in Fig.~\ref{figure6}. We choose $\gamma=0.1\omega_{1}$ only for quickly getting the main results. That is to say, the influence of the decay can be safely disregarded during the charging period, and considering the system of the two LC circuits as a closed system is a good theoretical approximation. The conclusions for the closed system drawn in the previous section are of significance.

\subsection{D.~Experimental consideration of our model}
The structures of the two parallel plate capacitors are assumed to be identical. The plates are squares with the side length of $1$mm. The distance between the two plates of each capacitor is $90\mu$m. The relative positions of the two capacitors are shown in Fig.~\ref{figure1}. The four plates are located in parallel to each other, and the vacuum is the dielectric. The active area is $1$mm$^{2}$. If the distance between the two capacitors in the nearest plates is $20\mu$m, we can obtain $k_{C}=-0.201$ according to the method given in Ref.~\cite{L.Huang2015}. The result ignores the edge effect of the capacitors. It is reasonable for the given parameters. For the two coupled inductive coils, we only make an estimate that the number of coil turns is very small, assuming that the radiuses of the coils are $1$mm and all of the coils are coaxial. The nearest ends of the two inductive coils are apart $50\mu$m (only quite $5\%$ of coil radius). Generally, $k_{L}$ must be at the range of $(-1, 1)$, Max$[E_{e}(t)]$ and Max$[R(t)]$ can be given in Fig.~\ref{figure4}. If the edge effect is ignored and there is basically no leakage phenomenon, $k_{L}$ will be close to $1$. By further numerical verification, Max$[E_{e}(t)]=82.12\hbar\omega_{1}$ and Max$[R(t)]=0.892$ for $k_{L}=0.99$ and $k_{C}=-0.201$. According to this analysis, it can be seen that the coupling capacitance $k_{C}$ is the core factor that determines the charging distance, and the wireless charging distance of the continuous variable quantum battery exceeds that of the qubit battery. This is the last feature of our continuous variable quantum battery.

\section{IV.~Conclusions}
In this paper, the wireless and remote charging of a continuous variable quantum battery based on the two coupled LC circuits is investigated. In the quibit battery, the characteristic frequencies of the charger and the battery must reach resonance for the battery to achieve maximum ergotropy. In our continuous variable quantum battery, the system can obtain more ergotropy only when the battery's frequency is greater than that of the charger. Ours can be called as a charging scheme of non-resonant continuous variable quantum battery. The primary advantage of the scheme is that it allows us to regulate the strengthes of the rotating and counter-rotating wave couplings by adjusting the relative position and distance between the capacitors and inductors, thereby providing an opportunity to study the roles of the couplings in battery charging dynamics. It is shown that the rotating and counter-rotating one play different roles during the charging process. The rotating wave coupling is helpful to transfer energy between the charger and the battery. The counter-rotating wave coupling can convert the interaction energy into ergotropy. Enhancing the counter-rotating wave coupling results in more ergotropy. For the case that the initial quantum state of the charger is a thermal state, the rotating or counter-rotating wave coupling alone can not make the ergotropy exist in the battery. It is necessary to make the two couplings cooperate with each other and produce quantum coherence in the system. The coherence is more significant than the quantum entanglement for the ergotropy of the continuous variable quantum battery. The charging distance of our quantum battery with wireless and remote charging far exceeds that of existing quibit batteries. Our work provides an general solution for the utilizing of the energy of the thermal reservoir.

\begin{acknowledgments}
\textbf{Acknowledgements}: J.W is supported by Research Foundation of Sichuan Minzu College (Grant No.~KYQD202402C).
P.P and G.Q.L acknowledge the supports from NSF of China (Grant No. 11405100) and the Natural Science Basic Research Plan in Shaanxi Province (Grant Nos.~2019JM-332 and 2020JM-507).
\end{acknowledgments}

\onecolumngrid

\newpage \clearpage
\onecolumngrid
\appendix

\section*{Supplementary Material for ``A continuous variable quantum battery with wireless and remote charging"}
\setcounter{table}{0} \renewcommand{\thetable}{S\arabic{table}} %
\setcounter{figure}{0} \renewcommand{\thefigure}{S\arabic{figure}} %
\setcounter{equation}{0} \def\theequation{S\arabic{equation}}

The density matrix of the continuous variable system has an infinite number of rows and columns, which is detrimental to the study of the properties of the system. The system can also be described by the Wigner functions or characteristic functions, which are complex numbers that correspond one-to-one to the density matrix. Once the characteristic functions are obtained, all the information about the system can be determined. Reference~\cite{Weedbrook2012} provides a method for calculating the characteristic function of the Gaussian states. For obtaining the characteristic function of the continuous variable quantum batteries in the main text, the expectation values of the operators $\langle\hat{a}_{2}^{\dag}\hat{a}_{2}\rangle$, $\langle\hat{a}_{2}^{\dag}\hat{a}_{2}^{\dag}\rangle$, $\langle\hat{a}_{2}\hat{a}_{2}\rangle$ and $\langle\hat{a}_{2}^{\dag}(t)\pm\hat{a}_{2}(t)\rangle$ need to be calculated firstly.

The expectation value of the operator does not vary due to the differences in the selected pictures (such as Schr$\ddot{o}$dinger's picture, Heisenberg's picture, and Dirac's picture). The Heisenberg's picture is chosen for our closed system. The evolution of the operators with time can be calculated analytically by the Laplace transform. The Hamiltonian is
\begin{eqnarray}
H &=&\hbar\omega _{1}\hat{a}_{1}^{\dag}\hat{a}_{1}+\hbar\omega _{2}\hat{a}_{2}^{\dag}\hat{a}_{2}+
g\hbar (\hat{a}_{1}^{\dag}+\hat{a}_{1})(\hat{a}_{2}^{\dag}+\hat{a}_{2})
+G\hbar (\hat{a}_{1}^{\dag}-\hat{a}_{1})(\hat{a}_{2}^{\dag}-\hat{a}_{2}).   \label{(S1)}
\end{eqnarray}
According to the Heisenberg equation of the operator, we can obtain the dynamical equations from Eq.~(\ref{(S1)}) if the role of the
environment has not been taken into account $(\gamma=0)$:
\begin{eqnarray}
\frac{d\hat{a}_{1}^{\dag}(t)}{dt} &=&i\omega
_{1}\hat{a}_{1}^{\dag}(t)+i(g-G)\hat{a}_{2}^{\dag}(t)+i(g+G)\hat{a}_{2}(t),  \notag \\
\frac{d\hat{a}_{1}(t)}{dt} &=&-i\omega
_{1}\hat{a}_{1}(t)-i(g+G)\hat{a}_{2}^{\dag}(t)+i(G-g)\hat{a}_{2}(t),  \notag \\
\frac{d\hat{a}_{2}^{\dag}(t)}{dt} &=&i\omega
_{2}\hat{a}_{2}^{\dag}(t)+i(g-G)\hat{a}_{1}^{\dag}(t)+i(g+G)\hat{a}_{1}(t),  \notag \\
\frac{d\hat{a}_{2}(t)}{dt} &=&-i\omega
_{2}\hat{a}_{2}(t)-i(g+G)\hat{a}_{1}^{\dag}(t)+i(G-g)\hat{a}_{1}(t).  \label{(S2)}
\end{eqnarray}

The variation of the operators with time in the above equations can be solved by the Laplace transform and its inverse transform. Through performing the Laplace transform for Eqs.~(\ref{(S2)}), we can obtain
\begin{eqnarray}
p\mathcal{L}[\hat{a}_{1}^{\dag}(t)]-\hat{a}_{1}^{\dag}(0) &=&i\omega
_{1}\mathcal{L}[\hat{a}_{1}^{\dag}(t)]+i(g-G)\mathcal{L}[\hat{a}_{2}^{\dag}(t)]+i(g+G)\mathcal{L}[\hat{a}_{2}(t)],  \notag \\
p\mathcal{L}[\hat{a}_{1}(t)]-\hat{a}_{1}(0) &=&-i\omega
_{1}\mathcal{L}[\hat{a}_{1}(t)]-i(g+G)\mathcal{L}[\hat{a}_{2}^{\dag}(t)]+i(G-g)\mathcal{L}[\hat{a}_{2}(t)],  \notag \\
p\mathcal{L}[\hat{a}_{2}^{\dag}(t)]-\hat{a}_{2}^{\dag}(0) &=&i\omega
_{2}\mathcal{L}[\hat{a}_{2}^{\dag}(t)]+i(g-G)\mathcal{L}[\hat{a}_{1}^{\dag}(t)]+i(g+G)\mathcal{L}[\hat{a}_{1}(t)],  \notag \\
p\mathcal{L}[\hat{a}_{2}(t)]-\hat{a}_{2}(0) &=&-i\omega
_{2}\mathcal{L}[\hat{a}_{2}(t)]-i(g+G)\mathcal{L}[\hat{a}_{1}^{\dag}(t)\}+i(G-g)\mathcal{L}[\hat{a}_{1}(t)].  \label{(S3)}
\end{eqnarray}
The Laplace transform is defined as $\mathcal{L}[f(t)]=\int_{0}^{+\infty }f(t)e^{-pt}dt$ for function $f(t)$. The Laplace transform of $\hat{a}_{1}^{+}(t)$, $\hat{a}_{1}(t)$, $\hat{a}_{2}^{+}(t)$ and $\hat{a}_{2}(t)$ can be obtained from Eqs.~(\ref{(S3)}). From the inverse Laplace transform, we can obtain
\begin{eqnarray}
\hat{a}_{1}(t)
&=&q_{1}(t)\hat{a}_{1}^{\dag}(0)+q_{2}(t)\hat{a}_{1}(0)+q_{3}(t)\hat{a}_{2}^{\dag}(0)+q_{4}(t)\hat{a}_{2}(0),
\notag \\
\hat{a}_{2}(t)
&=&f_{1}(t)\hat{a}_{1}^{\dag}(0)+f_{2}(t)\hat{a}_{1}(0)+f_{3}(t)\hat{a}_{2}^{\dag}(0)+f_{4}(t)\hat{a}_{2}(0).
\label{(S4)}
\end{eqnarray}
The conditions that the functions $f_{j}(t)$ and $q_{j}(t)$ $(j=1, 2, 3, 4)$ need to satisfy are
\begin{eqnarray}
\mathcal{L}[q_{1}(t)] &=&\frac{1}{Z}[2i(g^{2}-G^{2})\omega _{2}],  \notag \\
\mathcal{L}[q_{2}(t)] &=&\frac{1}{Z}\{p^{3}-i\omega _{1}p^{2}+(\omega
_{2}^{2}-4Gg)p+i[2\omega _{2}(G^{2}+g^{2})-\omega _{1}\omega _{2}^{2}]\},  \notag \\
\mathcal{L}[q_{3}(t)] &=&\frac{1}{Z}[-i(G+g)p^{2}+(G+g)(\omega _{2}-\omega
_{1})p+i(G+g)(4Gg-\omega _{1}\omega _{2})],  \notag \\
\mathcal{L}[q_{4}(t)] &=&\frac{1}{Z}[-i(g-G)p^{2}-(g-G)(\omega _{2}+\omega
_{1})p+i(g-G)(4Gg+\omega _{1}\omega _{2})],  \notag \\
\mathcal{L}[f_{1}(t)] &=&\frac{1}{Z}[-i(G+g)p^{2}+(G+g)(\omega _{1}-\omega
_{2})p+i(G+g)(4Gg-\omega _{1}\omega _{2})],  \notag \\
\mathcal{L}[f_{2}(t)] &=&\frac{1}{Z}[-i(g-G)p^{2}-(g-G)(\omega _{2}+\omega
_{1})p+i(g-G)(4Gg+\omega _{1}\omega _{2})],  \notag \\
\mathcal{L}[f_{3}(t)] &=&\frac{1}{Z}[2i(g^{2}-G^{2})\omega _{1}],  \notag \\
\mathcal{L}[f_{4}(t)] &=&\frac{1}{Z}\{p^{3}-i\omega _{2}p^{2}+(\omega
_{1}^{2}-4Gg)p+i[2\omega _{1}(G^{2}+g^{2})-\omega _{2}\omega _{1}^{2}]\}, \label{(S5)}
\end{eqnarray}
with $Z=p^{4}+(\omega _{1}^{2}+\omega _{2}^{2}-8Gg)p^{2}+[16G^{2}g^{2}-4\omega_{1}\omega _{2}(G^{2}+g^{2})+\omega _{1}^{2}\omega _{2}^{2}]$.
Based on the residue theorem, one can solve $Z=0$ with $p$ and find out the corresponding pole point. The covariance matrix of the Gaussian states and the expectation values of $\hat{a}_{2}^{\dag}(t)\pm \hat{a}_{2}(t)$ can be solved by Eqs.~(\ref{(S4)}). Therefore, the dynamics of the closed system can be determined.

For the open system $(\gamma\neq0)$, it is convenient to use the master equations. The variation of the expectation value of the operator $\langle\widehat{O}\rangle$ with time can be calculated by $\frac{d\langle\widehat{O}\rangle }{dt}=\mathrm{Tr}\left[\widehat{O}\frac{d\hat{\varrho}(t)}{dt}\right]$, where $\widehat{O}$ is an operator that does not obviously contain time. If we take the environment into account, the expectation values of the operators $\langle\hat{a}_{m}^{\dag}\hat{a}_{n}^{\dag}\rangle, \langle \hat{a}_{m}^{\dag}\hat{a}_{n}\rangle$ and $\langle \hat{a}_{m}\hat{a}_{n}\rangle$ ($m,n=1$ or $2$) are governed by the following ordinary differential equations:
\begin{eqnarray}
\frac{d\langle \hat{a}_{1}^{\dag}\hat{a}_{1}\rangle }{dt} &=&-i(G+g)(%
\langle \hat{a}_{1}^{\dag}\hat{a}_{2}^{\dag}\rangle -\langle
\hat{a}_{1}\hat{a}_{2}\rangle )-i(g-G)(\langle \hat{a}_{1}^{\dag}\hat{a}_{2}\rangle
-\langle \hat{a}_{1}\hat{a}_{2}^{\dag}\rangle )-\gamma\langle
\hat{a}_{1}^{\dag}\hat{a}_{1}\rangle +\gamma n_{th},  \notag \\
\frac{d\langle \hat{a}_{2}^{\dag}\hat{a}_{2}\rangle }{dt} &=&-i(G+g)(%
\langle \hat{a}_{1}^{\dag}\hat{a}_{2}^{\dag}\rangle -\langle
\hat{a}_{1}\hat{a}_{2}\rangle )-i(G-g)(\langle \hat{a}_{1}^{\dag}\hat{a}_{2}\rangle
-\langle \hat{a}_{1}\hat{a}_{2}^{\dag}\rangle ), \notag \\
\frac{d\langle \hat{a}_{1}^{\dag}\hat{a}_{2}^{\dag}\rangle }{dt} &=&i(\omega
_{1}+\omega _{2})\langle \hat{a}_{1}^{\dag}\hat{a}_{2}^{\dag}\rangle
+i(G+g)(\langle \hat{a}_{1}^{\dag}\hat{a}_{1}\rangle +\langle
\hat{a}_{2}^{\dag}\hat{a}_{2}\rangle +1)+i(g-G)(\langle \hat{a}_{1}^{\dag2}\rangle
+\langle \hat{a}_{2}^{\dag2}\rangle )-\gamma /2\langle
\hat{a}_{1}^{\dag}\hat{a}_{2}^{\dag}\rangle,  \notag \\
\frac{d\langle \hat{a}_{1}\hat{a}_{2}\rangle }{dt} &=&-i(\omega _{1}+\omega
_{2})\langle \hat{a}_{1}\hat{a}_{2}\rangle -i(G+g)(\langle
\hat{a}_{1}^{\dag}\hat{a}_{1}\rangle +\langle \hat{a}_{2}^{\dag}\hat{a}_{2}\rangle
+1)-i(g-G)(\langle \hat{a}_{1}^{2}\rangle +\langle
\hat{a}_{2}^{2}\rangle )-\gamma /2\langle \hat{a}_{1}\hat{a}_{2}\rangle,
\notag \\
\frac{d\langle \hat{a}_{1}^{\dag}\hat{a}_{2}\rangle }{dt} &=&i(\omega
_{1}-\omega _{2})\langle \hat{a}_{1}^{\dag}\hat{a}_{2}\rangle
-i(G+g)(\langle \hat{a}_{1}^{\dag2}\rangle -\langle
\hat{a}_{2}^{2}\rangle )-i(g-G)(\langle \hat{a}_{1}^{\dag}\hat{a}_{1}\rangle
-\langle \hat{a}_{2}^{\dag}\hat{a}_{2}\rangle )-\gamma /2\langle
\hat{a}_{1}^{\dag}\hat{a}_{2}\rangle,  \notag \\
\frac{d\langle \hat{a}_{1}\hat{a}_{2}^{\dag}\rangle }{dt} &=&i(\omega
_{2}-\omega _{1})\langle \hat{a}_{1}\hat{a}_{2}^{\dag}\rangle
-i(G+g)(\langle \hat{a}_{2}^{\dag2}\rangle -\langle
\hat{a}_{1}^{2}\rangle )+i(g-G)(\langle \hat{a}_{1}^{\dag}\hat{a}_{1}\rangle
-\langle \hat{a}_{2}^{\dag}\hat{a}_{2}\rangle )-\gamma /2\langle
\hat{a}_{1}\hat{a}_{2}^{\dag}\rangle,  \notag \\
\frac{d\langle \hat{a}_{1}^{2}\rangle }{dt} &=&-2i\omega
_{1}\langle \hat{a}_{1}^{2}\rangle -2i(G+g)\langle
\hat{a}_{1}\hat{a}_{2}^{\dag}\rangle -2i(g-G)\langle \hat{a}_{1}\hat{a}_{2}\rangle
-\gamma\langle \hat{a}_{1}^{2}\rangle,  \notag \\
\frac{d\langle \hat{a}_{1}^{\dag2}\rangle }{dt} &=&2i\omega
_{1}\langle \hat{a}_{1}^{\dag2}\rangle +2i(G+g)\langle
\hat{a}_{1}^{\dag}\hat{a}_{2}\rangle +2i(g-G)\langle
\hat{a}_{1}^{\dag}\hat{a}_{2}^{\dag}\rangle -\gamma\langle \hat{a}_{1}^{\dag2}\rangle,
\notag \\
\frac{d\langle \hat{a}_{2}^{2}\rangle }{dt} &=&-2i\omega
_{2}\langle \hat{a}_{2}^{2}\rangle -2i(G+g)\langle
\hat{a}_{1}^{\dag}\hat{a}_{2}\rangle -2i(g-G)\langle \hat{a}_{1}\hat{a}_{2}\rangle,
\notag \\
\frac{d\langle \hat{a}_{2}^{\dag2}\rangle }{dt} &=&2i\omega
_{2}\langle \hat{a}_{2}^{\dag2}\rangle +2i(G+g)\langle
\hat{a}_{1}\hat{a}_{2}^{\dag}\rangle +2i(g-G)\langle
\hat{a}_{1}^{\dag}\hat{a}_{2}^{\dag}\rangle.  \label{(S6)}
\end{eqnarray}
The evolution of $\langle\hat{a}_{m}^{\dag}\hat{a}_{n}^{\dag}\rangle$, $\langle \hat{a}_{m}^{\dag}\hat{a}_{n}\rangle$ and $\langle \hat{a}_{m}\hat{a}_{n}\rangle$ for the open system can be obtained by solving the system with $10$ ordinary differential equations (ODEs). The change of the expectation values of the operators $\langle\hat{a}_{2}^{\dag}\hat{a}_{2}\rangle$, $\langle\hat{a}_{2}^{\dag}\hat{a}_{2}^{\dag}\rangle$ and $\langle\hat{a}_{2}\hat{a}_{2}\rangle$ with time can be calculated from Eqs.~(\ref{(S6)}). The expectation values of $\langle\hat{a}_{2}^{\dag}(t)\pm\hat{a}_{2}(t)\rangle$ can be obtained by solving the following equations
\begin{eqnarray}
\frac{d\langle \hat{a}_{1}\rangle }{dt} &=&-i\omega _{1}\langle
\hat{a}_{1}\rangle -ig(\langle \hat{a}_{2}^{\dag}\rangle +\langle
\hat{a}_{2}\rangle )-iG(\langle \hat{a}_{2}^{\dag}\rangle -\langle
\hat{a}_{2}\rangle )-\gamma /2\langle \hat{a}_{1}\rangle,  \notag \\
\frac{d\langle \hat{a}_{1}^{\dag}\rangle }{dt} &=&i\omega _{1}\langle
\hat{a}_{1}^{\dag}\rangle +ig(\langle \hat{a}_{2}^{\dag}\rangle +\langle
\hat{a}_{2}\rangle )-iG(\langle \hat{a}_{2}^{\dag}\rangle -\langle
\hat{a}_{2}\rangle )-\gamma /2\langle \hat{a}_{1}^{\dag}\rangle,  \notag \\
\frac{d\langle \hat{a}_{2}\rangle }{dt} &=&-i\omega _{2}\langle
\hat{a}_{2}\rangle -ig(\langle \hat{a}_{1}^{\dag}\rangle +\langle
\hat{a}_{1}\rangle )-iG(\langle \hat{a}_{1}^{\dag}\rangle -\langle
\hat{a}_{1}\rangle ),  \notag \\
\frac{d\langle \hat{a}_{2}^{\dag}\rangle }{dt} &=&i\omega _{2}\langle
\hat{a}_{2}^{\dag}\rangle +ig(\langle \hat{a}_{1}^{\dag}\rangle +\langle
\hat{a}_{1}\rangle )-iG(\langle \hat{a}_{1}^{\dag}\rangle -\langle
\hat{a}_{1}\rangle ).  \label{(S7)}
\end{eqnarray}
Based on such calculations, the characteristic functions of the Gaussian states for the open system can be determined. It is noted that the evolution of the above expectation values for the closed system can also be investigated by the master equations under the condition $\gamma=0$. The results are consistent with that given by the method of the Laplace transform.

\end{document}